\newcommand{\CLASSINPUTtoptextmargin}{1.7cm}
\newcommand{\CLASSINPUTbottomtextmargin}{1.7cm}
\newcommand{\CLASSINPUTinnersidemargin}{1.5cm}
\newcommand{\CLASSINPUToutersidemargin}{1.5cm}
\documentclass[conference]{IEEEtran}
\IEEEoverridecommandlockouts
\usepackage{cite}
\usepackage{amsmath,amssymb,amsfonts}
\usepackage{algorithmic}
\usepackage{graphicx}
\usepackage{textcomp}
\usepackage{xcolor}
\usepackage{setspace}
\usepackage{url}
\usepackage{caption}

\def\BibTeX{{\rm B\kern-.05em{\sc i\kern-.025em b}\kern-.08em
    T\kern-.1667em\lower.7ex\hbox{E}\kern-.125emX}}
\begin{document}

\title{Perceptual impact of the loss function on deep-learning image coding performance\\
 \thanks{
This work was supported by the Fundação para a Ciência e a Tecnologia (FCT) through the Project entitled Deep compression: emerging pAradigm foR Image codiNG under Grant PTDC/EEI-COM/7775/2020.}
}

\author{
    \IEEEauthorblockN{Shima Mohammadi\IEEEauthorrefmark{1}, Jo\~{a}o Ascenso\IEEEauthorrefmark{2}}
    \IEEEauthorblockA{\IEEEauthorrefmark{1}\IEEEauthorrefmark{2}Instituto Superior Técnico - Instituto de Telecomunicações}
    \IEEEauthorblockA{\IEEEauthorrefmark{1}shima.mohammadi@lx.it.pt, \IEEEauthorrefmark{2}joao.ascenso@lx.it.pt}
    \vspace{-30pt}
}

\maketitle
\begin{abstract}
 Nowadays, deep-learning image coding solutions have shown similar or better compression efficiency than conventional solutions based on hand-crafted transforms and spatial prediction techniques. These deep-learning codecs require a large training set of images and a training methodology to obtain a suitable model (set of parameters) for efficient compression. The training is performed with an optimization algorithm which provides a way to minimize the loss function. Therefore, the loss function plays a key role in the overall performance and includes a differentiable quality metric that attempts to mimic human perception. The main objective of this paper is to study the perceptual impact of several image quality metrics that can be used in the loss function of the training process, through a crowdsourcing subjective image quality assessment study. From this study, it is possible to conclude that the choice of the quality metric is critical for the perceptual performance of the deep-learning codec and that can vary depending on the image content. 
\end{abstract}
\begin{IEEEkeywords}
Learning-based image compression, percepual optimization, loss function, image quality metrics
\end{IEEEkeywords}
\vspace{-4pt}

\section{Introduction}

Nowadays, deep neural (DL) networks are an efficient way to compress visual content, leveraging on the availability of large image datasets and highly parallelizable graphic processing units (GPUs). For some image-related applications, such as cloud storage, visual surveillance, image collection and storage, there are compelling advantages since higher compression efficiency can lead to significant savings in storage or transmission costs. Moreover, image processing and computer vision tasks can be performed with lower complexity or even high accuracy by using as input the compressed domain representation (and thus artifact free) instead of the original or decoded images \cite{JPEG-WhitePaper}.
\\Recently, JPEG has initiated a new initiative to standardize the first image coding standard based on machine learning. The aim is to \textit{offer a single stream, compact compressed domain representation, targeting both human visualization with significant compression efficiency improvement over image coding standards in common use at equivalent subjective quality and effective performance for image processing and computer vision tasks}\cite{JPEG-WhitePaper}. Other key requirements include hardware/software implementation-friendly encoding and decoding, support for 8 and 10-bit depth, efficient coding of images with text and graphics and progressive decoding.
\\The pipeline for many DL based image compression solutions is still similar to conventional image codecs, the original image is first transformed to obtain a compact (latent) representation, which is then quantized/rounded, followed by entropy coding to exploit any statistical dependencies. In conventional image compression, a linear hand-crafted transformation (e.g. DCT or DWT) is used, and all the three main components are optimized separately. On the other hand, DL codecs use a cascade set of neural network layers (some performing non-linear operations) as the transform but also to obtain accurate probability models for the entropy coding engine. Usually, all parts of the DL image compression pipeline are jointly optimized in an end-to-end fashion. A DL image compression is trained by iteratively processing the data through the network (feedforward) and then by backpropagating the loss to compute the model parameters until no further improvements are achieved. The loss function typically includes the bitrate and the score given by a quality metric that computes the error between the decoded and the original image. Therefore, objective image quality assessment is one of the most important aspects of the training procedure and has a very significant impact on the obtained model parameters and thus on the artifacts that may appear on the final decoded image.
\\ Often, image coding solutions are trained using the mean squared error (MSE), a popular fidelity metric, which may have a poor correlation with quality perception of humans or the MS-SSIM, which may lead to artifacts such as color or contrast shifts or even cartoon-like regions \cite{ascenso2020learning}. Moreover, there are many other objective quality metrics available in the literature that can be used, and is not known which one leads to decoded images which have a higher preference among humans. In this context, the objective of this paper is to study the impact of using several image quality metrics in the training methodology of DL based image codecs and thus find which metric results in higher perceived quality and for how much. The main novel contribution is the subjective assessment campaign which evaluates several DL image codecs which only differ on the way their respective models were obtained, namely in the objective quality metric of the loss function. The experimental study presented in this paper will allow to understand the limitations and gains of the several quality assessment metrics available nowadays for  training and conclude about the quality metric that should be used for the optimization of DL image coding solutions, such as the ones that are now being considered for image coding standardization. 
\\The rest of the paper is organized as following: Section \ref{relat_work} describes the related work while Section \ref{subj_assessment} describes the subjective evaluation methodology. Section \ref{performance_eval} presents the experimental results obtained and the key conclusions of this study. Finally, Section \ref{conclusions} concludes about the results of this work.
\vspace{-0.5mm}
\section{Related Work}\label{relat_work}

The performance of learning-based image compression solutions were evaluated in the past in several studies \cite{ascenso2020learning}\cite{upenik2021large}, both using objective quality metrics and subjective assessment procedures. For example, in \cite{ascenso2020learning}, several DL image coding based solutions have been evaluated in comparison to conventional image codecs such as HEVC, WebP, JPEG 2000 and JPEG, using a side-by-side Double Stimulus Impairment Scale subjective assessment protocol in a lab controlled environment. The results have shown that DL image coding solutions have competitive compression efficiency in comparison to conventional image codecs such as HEVC Intra. In \cite{upenik2021large}, more recent DL image coding solutions have been evaluated using a crowdsourcing based Double Stimulus Continuous Quality Scale methodology, which is able to account for cases where the decoded image quality may be higher than the reference image quality. The results have also shown that DL image coding approach can achieve promising compression performance.
\par
Moreover, in \cite{ding2021comparison}, several full reference image quality assessment metrics were evaluated regarding their use for four image processing tasks, namely denoising, deblurring, super-resolution and image compression. However, for image compression, bitrate is not considered in the loss function which severely impacts the conclusions taken from the experimental results. The work presented here is the first that evaluates several objective quality metrics when used for optimization for a widely popular image compression solution available from the literature, considering a bitrate constrain in the loss function.

\vspace{-0.5mm}
\section{Perceptual Optimization of DL Image Compression Solutions}

The key idea of this paper is to evaluate several popular image quality metrics when they are used to guide the training process. Therefore, the image codec architecture is kept fixed, this means all the hyperparameters, such as the number of layers and the number of channels, while several loss functions that only differ on the quality metric are used. The only constraint is that the quality metric can be differentiable but there are many than meet this requirement, especially those that exploit features obtained through convolutional layers, e.g. LPIPS\cite{zhang2018LPIPS}. 

The variational autoencoder with an hyperprior (VAE-Hyper) codec\cite{balle2018variational} is the selected deep-learning codec since it is rather popular and many other more recent deep-learning codecs (e.g. \cite{Cheng2020}) have the same structure with a set of convolutional layers and non-linear activations and the transmission of side information (obtained with a seperate encoder-decoder) that is used to obtain a probability model for the entropy coding of the latent representation.

In the VAE-Hyper codec, the loss, as defined in \ref{equ1}, represents the joint rate-distortion trade-off, where the rate (R) is the code length of quantized coefficients $\overline{Y}$, and the distortion $D$ is the difference between the original image $X$ and the reconstructed image $\overline{X}$. The rate-distortion trade-off is controlled by $\lambda$ which corresponds to the usual Lagrange multiplier in conventional image coding. 

\begin{equation}
Loss=\lambda(D(X-\overline{X}))+R(\overline{Y}))
\label{equ1}
\vspace{-0.1cm}
\end{equation}

To obtain different points on the rate-distortion curve, multiple models can be trained by varying the value of $\lambda$. Therefore, several VAE-Hyper codecs are obtained which only differ on the model but not on the architecture. Each VAE-Hyper codec was trained for several target bitrates corresponding to low, medium, and high bitrates, namely 0.1bpp, 0.4bpp and 0.7bpp, with 10\% of tolerance, by adjusting the $\lambda$ value. The three target bitrates correspond to low, medium, and high qualities, and were found based on the results of a viewing study made by a small group of experts.

\vspace{-4pt}
\section{Subjective Quality Assessment Methodology}\label{subj_assessment}

This section describes the crowdsourcing-based subjective test methodology that was used to evaluate the several distortion metrics under consideration.

\subsection{Subjective Evaluation Methodology}

To obtain reliable and accurate results about the user preference for each VAE-Hyper-X codec, the subjective test follows a ranking subjective test methodology. In this case, a pairwise comparison (PC) is used, where two images are shown side by side, and subjects are asked to select the image with the highest quality. The PC subjective assessment methodology has several advantages, namely not requiring training associated to the meaning of quality scale (also avoiding any misunderstanding), robustness to viewing conditions (which often occur in subjective tests) as well as high accuracy. The main disadvantage is that the number of pairs that subjects must evaluate increases exponentially with the number of stimuli, when all the possible pair combinations between all stimuli are used. However, several methods were previously proposed to overcome this shortcoming. In this study, the Swiss system (often employed in Chess Tournaments) was used to reduce the number of pairs under evaluation due to its popularity \cite{lin2015mcl}\cite{ponomarenko2009tid2008}. The objective of this approach is to create a sorted list of images according to its perceptual quality. First, adjacent images of an initial ordered list are compared by subjects, and are swapped according to the scores. This process is repeated for the same subject until no further swapping is required. The initial list was also defined by expert viewing by a small group of experts. In this work, the Swiss system was implemented in the crowdsourcing platform at the client (browser) side. \par

The PC test was divided into three different sessions to reduce subjects fatigue and thus meet ITU BT 500-13 recommendation of having test with a maximum of 30min duration. In each session, images obtained by coding and decoding with the several VAE-Hyper-X codecs are shown in a random order. In a pair, the two decoded images were always obtained from the same reference image and from the same target bitrate. 
\vspace{-6pt}
\subsection{Test Material}\label{test_material}

\begin{figure}[htbp]
\centerline{
\begin{tabular}{@{}c@{}}
    {\includegraphics[scale=0.07]{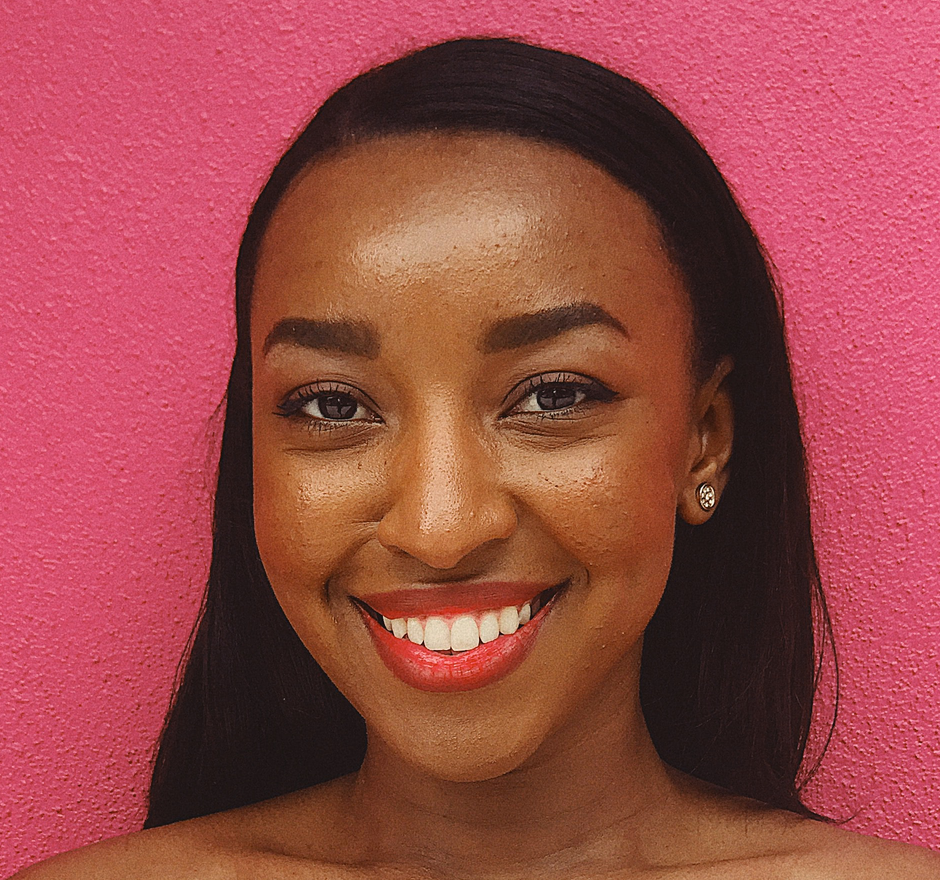}} \\ 
    \small(a) \textit{Woman}
  \end{tabular}
  \begin{tabular}{@{}c@{}}
    {\includegraphics[scale=0.07]{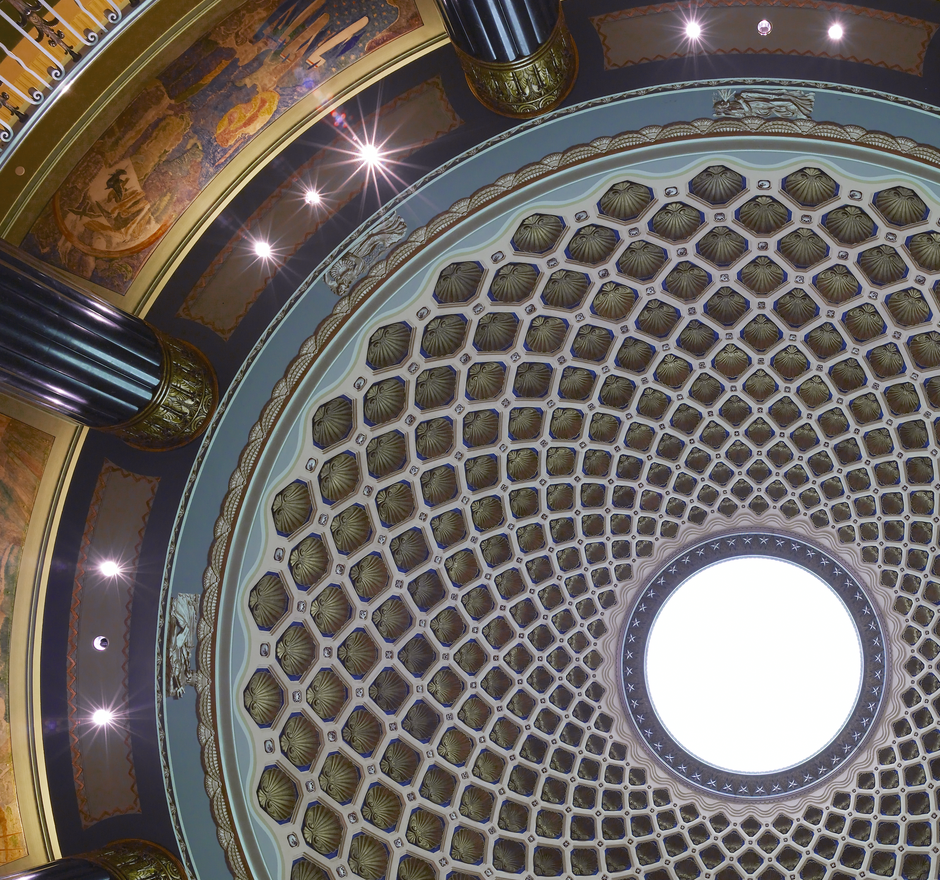}} \\ 
    \small (b) \textit{Rotunda Mosta}
  \end{tabular}
  \begin{tabular}{@{}c@{}}
    {\includegraphics[scale=0.07]{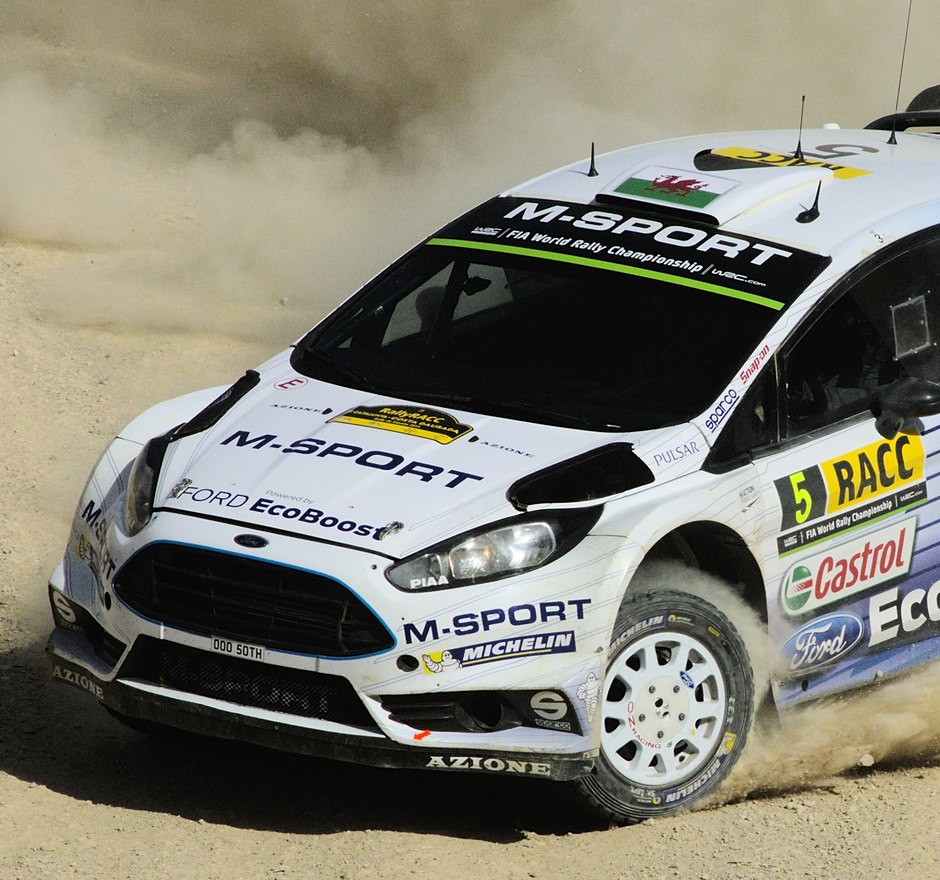}} \\ 
    \small (c) \textit{Racing car}
  \end{tabular}}
\centerline{
\begin{tabular}{@{}c@{}}
    {\includegraphics[scale=0.07]{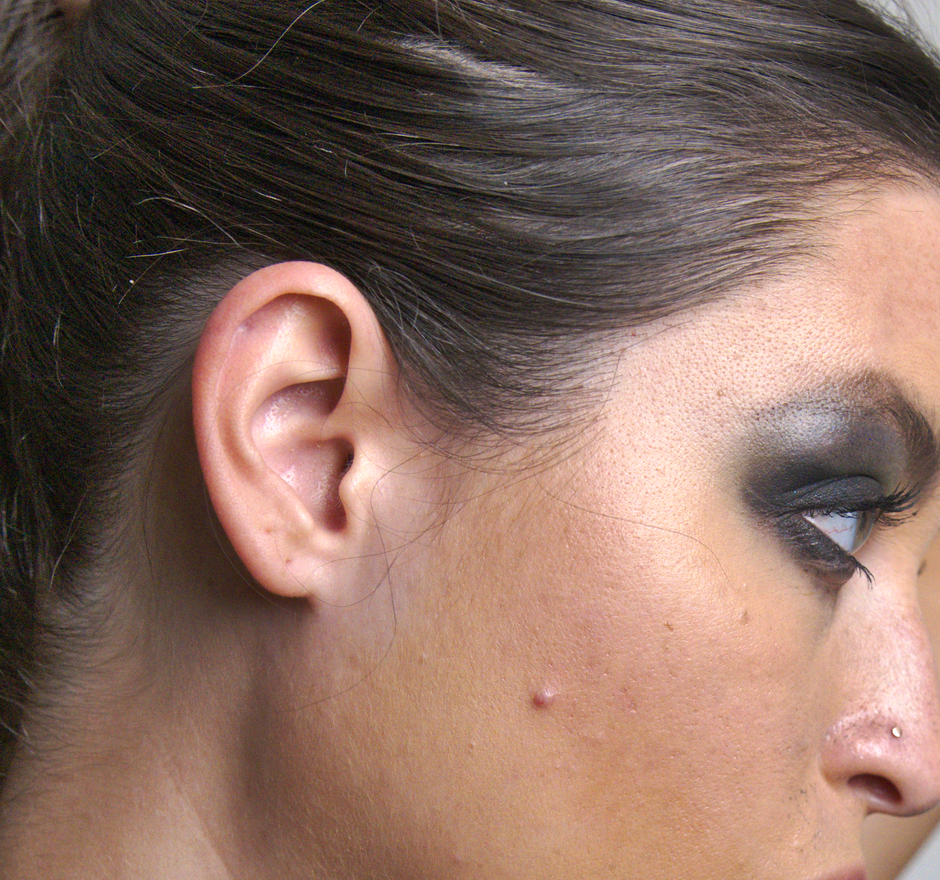}} \\
    \small (d) \textit{Ponytail}
  \end{tabular}
  \begin{tabular}{@{}c@{}}
    {\includegraphics[scale=0.07]{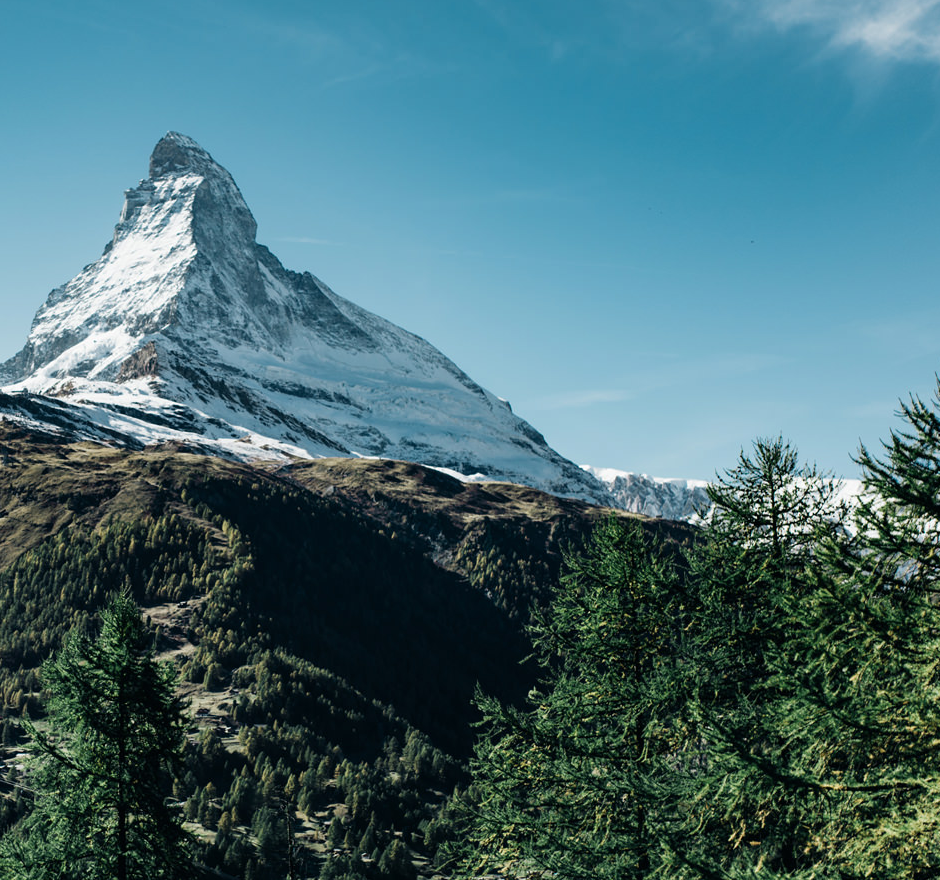}} \\
    \small (e) \textit{Matterhorn}
  \end{tabular}
  \begin{tabular}{@{}c@{}}
    {\includegraphics[scale=0.07]{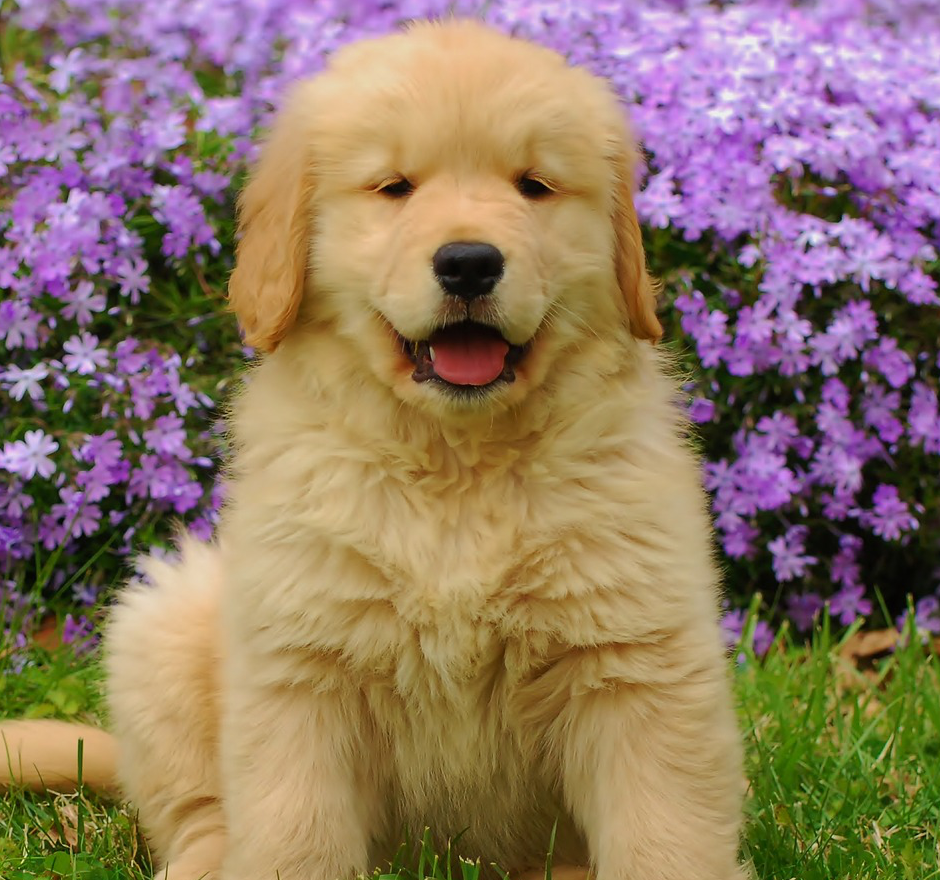}} \\
    \small (f) \textit{Dog}
  \end{tabular}}
\caption{Cropped test images used in the subjective study.}
\label{fig1}
\vspace{-12pt}
\end{figure}

To evaluate all the variants of the VAE-Hyper codec, six test images of JPEG AI dataset were selected. These images are selected to cover a large range of image content characteristics, including high and low texture regions. The coded images were cropped to fit the side-by-side layout of the crowdsourcing platform considering a minimum display resolution of $1920\times1080$. This resolution is enforced during the subjective test. Fig.\ref{fig1} shows the cropped images used in the subjective test. Also, the complete dataset is available online at the DL codec optimization study repository \cite{repository}.

\subsection{Training Procedure}

The VAE-Hyper training is done on patches of $256\times256$ size images of the JPEG AI training and validation dataset. The Pytorch codec implementation of the CompressAI library was used \cite{compressai}. The number of epochs was 200 and the learning rate ($lr$) is initially 1E-5, and when the validation loss is stable $lr$ is divided by two. There are two training steps: in the 1st step, the model is trained using MSE as loss function, whereas in the 2nd step, it is fine-tuned with the selected image quality metrics. This selection was based on an initial informal analysis of the VAE-Hyper performance for a large number of full-reference quality metrics. For example, it was found that some such as VIF\cite{VIF} and CW-SSIM \cite{CW-SSIM} have lead to rather low performance or training instability and thus were not considered. The final selection is shown in Table \ref{tab1}. The training procedure was performed on NVIDIA GTX-GeForce 1660 GPU.
\vspace{-7pt}
\begin{table}[htbp]
\footnotesize
\caption{Selected image quality metrics for the VAE-Hyper} 
\begin{center}
\vspace{-7pt}
\begin{tabular}[\textwidth]{|l|p{6.3cm}|}
\hline
\textbf{Metrics}&{\textbf{Short Description}} \\
\hline
MSE & Measures pixel-wise squared differences\\
\hline
SSIM\cite{wang2004image} & Measures the degradation in structural information  \\ 
\hline
MS-SSIM\cite{wang2003multiscale} & SSIM extension that supports variations in image resolution and viewing conditions \\
\hline
FSIM\cite{zhang2011fsim} & Exploits phase congruency and gradient information \\
\hline
GMSD\cite{xue2013gradient} & Measures pixel-wise gradient differences  \\ 
\hline
LPIPS & Measures similarity using deep features \\
\hline
DISTS\cite{ding2020image} & Measures structural distortions with a tolerance for texture resampling \\
\hline
NLPD\cite{laparra2016perceptual} & Measures root mean square error differences in normalized laplacian domain \\
\hline
VSI\cite{zhang2014vsi} & Exploits saliency features for local distortion computation  \\
\hline
\end{tabular}
\vspace{-0.4cm}
\label{tab1}
\end{center}
\end{table}

\subsection{Experimental Setup}

The subjective experiment software platform is a web framework that was designed to perform crowdsourcing-based pairwise subjective quality assessment \cite{MMarquesApp}. This web client-server application relies on JavaScript and a MongoDB database to collect and store all the information. The web interface is simple and usable with common web browsers. The subjects were recruited from Amazon Mechanical Turk (AMT), a crowdsourcing website that allows to hire remotely located crowdworkers to perform discrete on-demand tasks. 
\par
Subjects were invited to participate in the test through AMT without any restriction in terms of geographical location or age. The crowdsourcing subjective evaluation platform asks first for the e-mail address, name, age, display size, and gender of the subjects. To minimally control the viewing conditions, it was enforced a restriction to only allow subjects to participate with a display that has a minimum resolution of $1920\times 1080$, and display size of at least 13 inches. The images are never upsampled or downsampled and thus are shown in their natural resolution indepdently of the display resolution. 
\par
\begin{figure}[htbp]
\centerline{\includegraphics[scale=0.1]{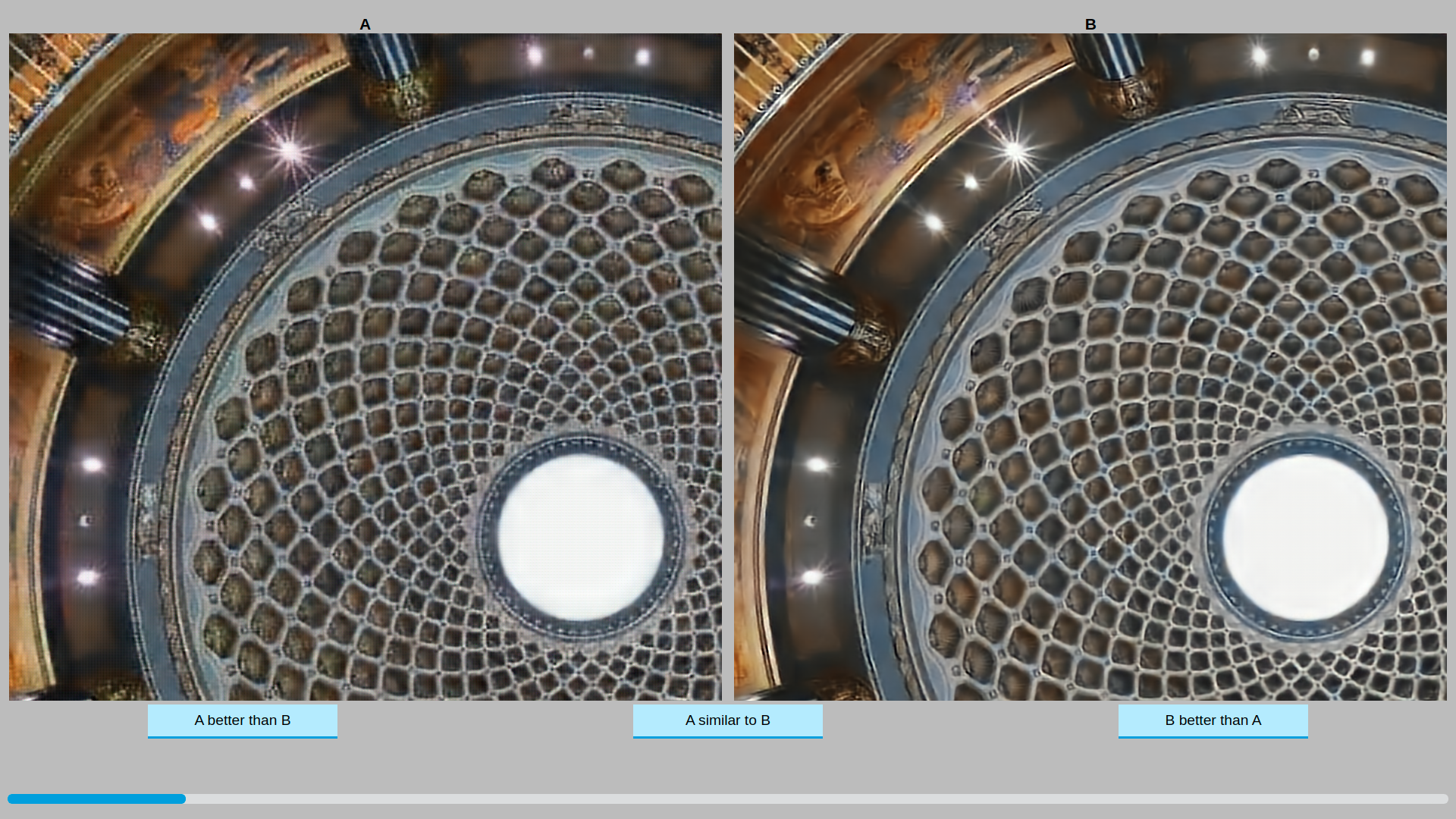}}
\setlength\abovecaptionskip{0pt}
\setlength\belowcaptionskip{0pt}
\caption{Layout of the web graphical user interface that implements the pairwise comparison subjective assessment test}
\label{fig2}
\vspace{-12pt}
\end{figure}

Furthermore, subjects had to perform a small training phase before conducting the actual test. The objective of this phase is to familiarize subjects with the platform interface and controls as well as the objective of the subjective test. The training phase consists in three pairs of images where the decision is rather obvious; these images are different from the the ones shown during the evaluation phase. During the training phase, subjects are informed about the correct decision, which they must select to advance, in order to familiarize them with the platform. Images were shown to the subjects, side by side and in random order with three questions: Image A better than image B, image B better than image A, and image A similar to image B. A layout of the platform is shown in Fig.\ref{fig2}.

\subsection{Subjective Data Processing}
To process the data obtained in the subjective assessment test, the following operations were done.
\subsubsection{Outlier Detection}
Due to the crowdsourcing nature of the subjective test, full control of the subjects actions is not possible and thus, outliers may appear due to different viewing conditions, human errors or other abnormal behaviour. To detect and remove outliers from the data, the number of transitivity cycles in the test is counted according to the subjects judgements. The procedure used was inspired by \cite{FalahandAscenso}. Let $A>B$ imply that image $A$ was preferred over image $B$, and $A=B$ indicate a tie between the two images. Outliers can be detected when $(A=B) \cap (B<C)$ and thus according to the Swiss system, $B$ and $C$ must be swapped, followed by a comparison of $A$ and $C$. If $(A=C)\:or\:(A>C)$, a transitivity cycle is detected. The ratio of transitivity cycles is calculated as in (\ref{equ2}), 
\begin{equation}
R=1-d/h,
\label{equ2}
\end{equation}
where $d$ is the number of transitivity cycles for a subject, and $h$ is the total number of transitivity cycles that occur, i.e. $(A=B) \cap (B<C)$.

\subsubsection{Quality Score Computation}

First, a pairwise comparison matrix $PCM^s[i,j]$ is computed after each subject $s$ ranks all images, from the final ordered list obtained. This sorted list reflects the subjects preference and thus the perceptual quality of each decoded image. $PCM[i,j]$ represents the number of times image $i$ is preferred over image $j$. Therefore, $PCM[i,j]$ is populated assuming that each decoded image of the ordered list is preferred over all of the images in the descending positions of the list. In case the subject had no preference between two images (i.e. it is a tie), 0.5 is used instead. After all subjects have performed the test, a group preference matrix $PCM[i,j]$ is computed by aggregating the preferences matrices $PCM^s[i,j]$ of all the subjects, obtaining the number of times image $i$ is preferred over image $j$.

\par
Different techniques are available to compute psychometric quality scores from the group preference matrix. The most widely used methods are winning frequency, Bradley Terry (BT)\cite{BT} and Thurstone-Moller model\cite{Thurstone1994ALO}. The scores obtained with all these methods are highly correlated with each other (Pearson correlation coefficient of 0.985) and thus, the winning frequency is used in this work, due to its simplicity and popularity. The winning frequency is computed for each each bitrate and reference image from the group preference matrix. It corresponds to the number of votes each metric has received divided by the total number of votes.
\vspace{-0.5mm}
\section{Performance Evaluation}\label{performance_eval}

The results of the subjective assessment study are presented and analyzed in this section.

\subsection{Experimental Results}

\begin{figure}[htbp]
\centerline{
\begin{tabular}{@{}c@{}}
    {\includegraphics[scale=0.3]{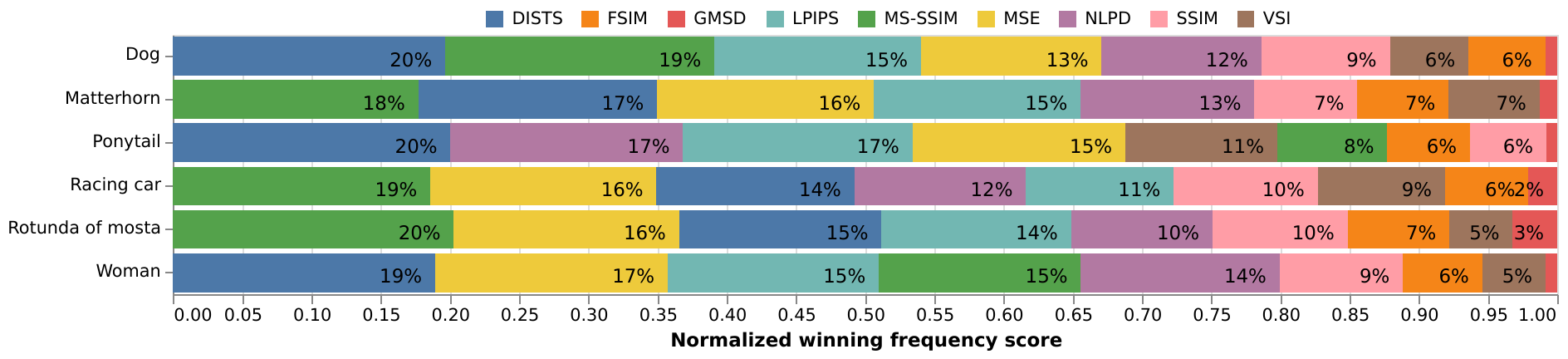}} \\
    \footnotesize (a) Target bitrate of 0.1bpp (Low bitrate)
  \end{tabular}}
\centerline{
\begin{tabular}{@{}c@{}}
    {\includegraphics[scale=0.3]{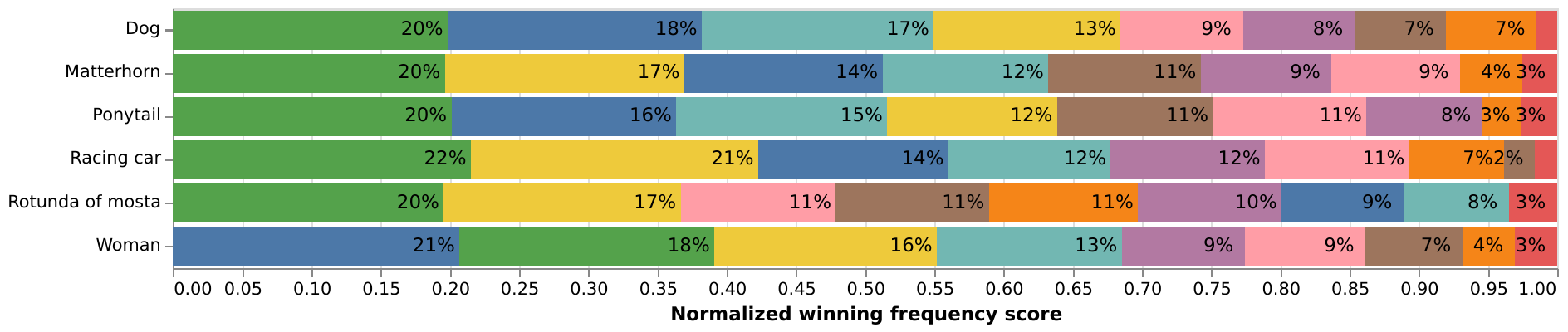}} \\
    \footnotesize (b) Target bitrate of 0.4bpp (Medium bitrate)
  \end{tabular}}
  
\centerline{
\begin{tabular}{@{}c@{}}
    {\includegraphics[scale=0.3]{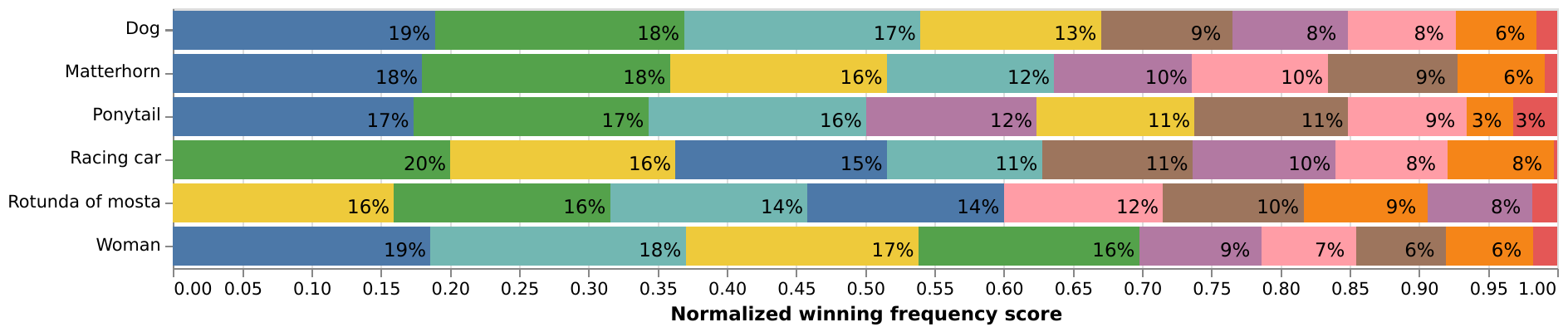}} \\
    \footnotesize (c) Target bitrate of 0.7bpp (High bitrate)
  \end{tabular}}

\caption{Normalized winning frequency results for the selected image quality metrics}  
\label{fig3}
\vspace{-12pt}
\end{figure}

The crowdsourcing subjective assessment campaign was performed by 120 users distributed equally in the three sessions. The subjects' age was distributed between 20 and 60 with an average of 34. Approximately 70\% of the subjects were male. The more common display resolution is $1920\times 1080$ and the more common display size is 15 inches. The number of outliers in the three sessions was 2, 4, 6, and the judgments made by these outliers were discarded from the data. Fig.\ref{fig3} shows the normalized winning frequency for each decoded image obtained by each VAE-Hyper codec. For better visualization, when the quality metric has a very low score (less than 2\%) the label is not shown in Fig.\ref{fig3}. The following conclusions can be taken: 
\begin{itemize}
\item DISTS, MS-SSIM are generally the best for all the three bitrates. Moreover, the 2nd best quality metric (often MSE) has a performance rather close to the 1st best performance quality metric.
\item Low bitrate (0.1bpp): DISTS and MS-SSIM have the best performance, with the exception of the \textit{Ponytail} image for which MS-SSIM ranks significantly lower. In this case, NLPD and LPIPS are the 2nd and the 3rd best.
\item Medium bitrates (0.4bpp): MS-SSIM has the best performance, outperforming other metrics 5 out of 6 times. However, DISTS has the highest performance for the \textit{Woman} image, which actually happens consistently for all the bitrates. 
\item High bitrate (0.7bpp): DISTS has the best overall performance while for two specific images, \textit{Racing Car} and \textit{Rotunda Mosta}, MS-SSIM and MSE provide better performance. Interestingly, MSE performs poorly for high bitrates than for low and medium bitrates.
\end{itemize}

\begin{figure*}[ht]
\setlength\abovecaptionskip{0pt}
\setlength\belowcaptionskip{0pt}
\centerline{
\begin{tabular}{@{}ccccc@{}}
    {\includegraphics[scale=0.25]{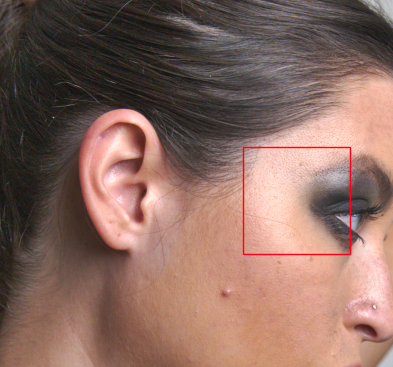}} \\
    \small (A) \textit{Ponytail}
  \end{tabular}
  
  \begin{tabular}{@{}c@{}}
    {\includegraphics[scale=0.25]{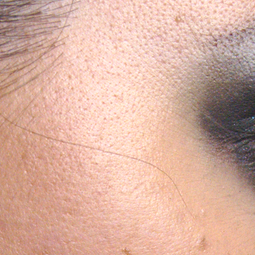}} \\
    \small (a) Cropped\\
      {\includegraphics[scale=0.25]{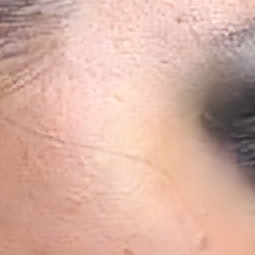}} \\
       \small (f) DISTS
  \end{tabular}
  
  \begin{tabular}{@{}c@{}}
    {\includegraphics[scale=0.25]{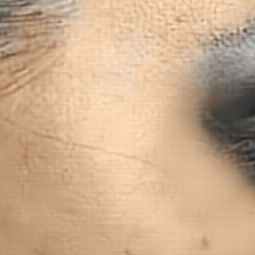}} \\
    \small (b) FSIM\\
      {\includegraphics[scale=0.25]{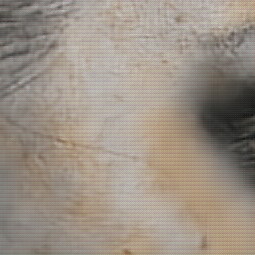}} \\
       \small (g) GMSD
  \end{tabular}
  
  \begin{tabular}{@{}c@{}}
    {\includegraphics[scale=0.25]{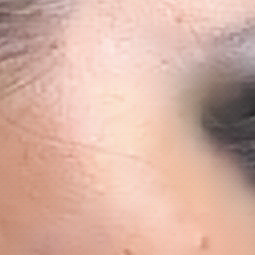}} \\
    \small (c) LPIPS\\
      {\includegraphics[scale=0.25]{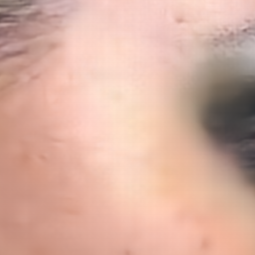}} \\
       \small (h) MSE
  \end{tabular}
  
  \begin{tabular}{@{}c@{}}
    {\includegraphics[scale=0.25]{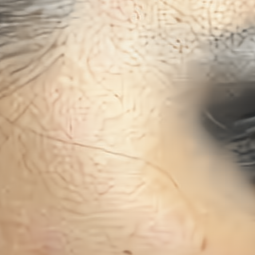}} \\
    \small (d) MS-SSIM\\
      {\includegraphics[scale=0.25]{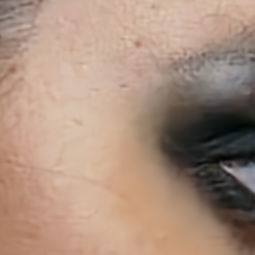}} \\
       \small (i) NLPD
  \end{tabular}
  
  \begin{tabular}{@{}cc@{}}
 
    {\includegraphics[scale=0.25]{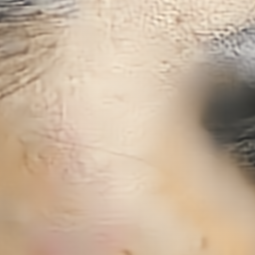}} \\
    \small (e) SSIM\\
     {\includegraphics[scale=0.25]{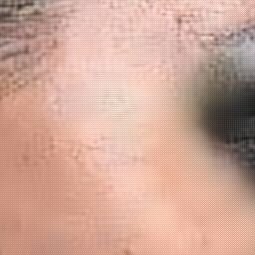}} \\
      \small (j) VSI
  \end{tabular}}
\label{fig4}
\vspace{-12pt}
\end{figure*}

\begin{figure*}[ht]
\setlength\abovecaptionskip{0pt}
\setlength\belowcaptionskip{0pt}
\centerline{
\begin{tabular}{@{}ccccc@{}}
   
    {\includegraphics[scale=0.25]{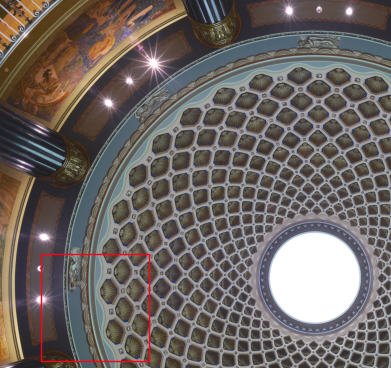}} \\
    \small (B) \textit{Rotunda Mosta}
  \end{tabular}
  
  \begin{tabular}{@{}c@{}}
    {\includegraphics[scale=0.25]{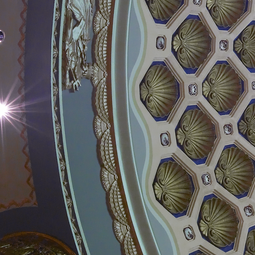}} \\
    \small (a) Cropped\\
      {\includegraphics[scale=0.25]{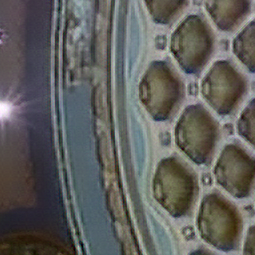}} \\
      \small (f) DISTS
  \end{tabular}
  
  \begin{tabular}{@{}c@{}}
    {\includegraphics[scale=0.25]{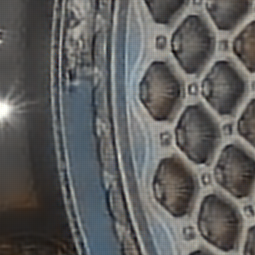}} \\
    \small (b) FSIM\\
      {\includegraphics[scale=0.25]{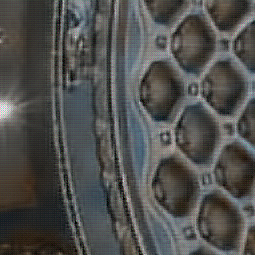}} \\
      \small (g) GMSD
  \end{tabular}
  
  \begin{tabular}{@{}c@{}}
    {\includegraphics[scale=0.25]{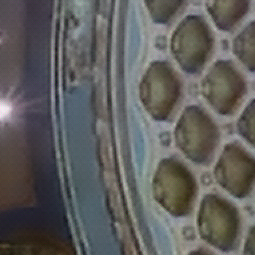}} \\
    \small (c) LPIPS\\
      {\includegraphics[scale=0.25]{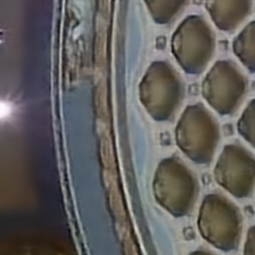}} \\
      \small (h) MSE
  \end{tabular}
  
  \begin{tabular}{@{}c@{}}
    {\includegraphics[scale=0.25]{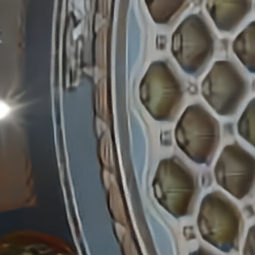}} \\
    \small (d) MS-SSIM\\
      {\includegraphics[scale=0.25]{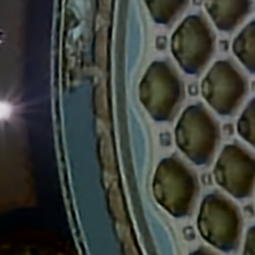}} \\
      \small (i) NLPD
  \end{tabular}
  
  \begin{tabular}{@{}cc@{}}
 
    {\includegraphics[scale=0.25]{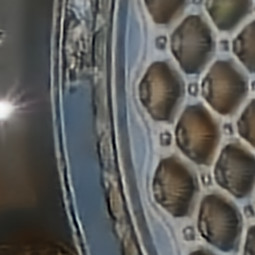}} \\
    \small (e) SSIM\\
     {\includegraphics[scale=0.25]{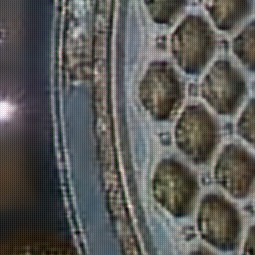}} \\
      \small (j) VSI
  \end{tabular}}
\vspace{+5pt}
\caption{Regions of the selected original and decoded images for VAE-Hyper for several image quality assessment metrics}
\label{fig5}
\vspace{-10pt}
\end{figure*}
\vspace{-0.5mm}
\subsection{Qualitative results}\label{Qualitative results}
The compressed images for two of the test images, \textit{Rotunda Mosta} and \textit{Ponytail} at low bitrate (0.1bpp) are shown in Fig.\ref{fig5}. 
\\Based on Fig.\ref{fig5}(A), The \textit{Ponytail} image optimized with MS-SSIM failed to generate natural colors for the skin of the face, which it was noticeable by the subjects. This is compatible with the previous results (Fig.\ref{fig3}), which shows that MS-SSIM metric preference is at 8\%, only slightly better than SSIM and FSIM scores. On the other hand, DISTS provides a high quality image even at lower bitrates, preserving important details.
\\In Fig.\ref{fig5}(B), MS-SSIM quality metric generates more sharp images, while DISTS and LPIPS much inferior results with the presence of ringing artifacts. This can also be confirmed by the previous results (Fig.\ref{fig3}), where MS-SSIM has the best performance.
\\As shown in Fig.\ref{fig5} in all the bitrates, NLPD and GMSD metric leads to color shifts and low contrast, and thus are not a good choice for image compression. As expected, MSE images lack detail and are somewhat blurry while not having noticeable artifacts for some images, such as \textit{Woman} and \textit{Rotunda Mosta}. 

\vspace{-6pt}
\section{Conclusions}\label{conclusions}
In this study the impact of the image quality metric used in the loss function of deep-learning image codecs was studied. A crowdsourcing based subjective image quality assessment study was performed using a pairwise ranking methodology. The conclusions show that the performance of the deep-learning image codec depends on the selected metric and based on the experimental results, DISTS and MS-SSIM offer the best performance overall. As future work, the performance impact of the adversarial loss (computed with a discriminative network as in generative adversarial networks) will be studied when used in combination with the the rate-distortion loss function presented here.
\vspace{-0.7mm}
\bibliographystyle{ieeetr}
\bibliography{citation}

\end{document}